\documentclass[usenatbib]{mn2e}

\usepackage{graphicx}

\title[Alfv{\'e}n waves in the solar wind]{Origin of long-period Alfv{\'e}n waves in the solar wind}
\author[T.V. Zaqarashvili and G. Belvedere]{T.V. Zaqarashvili$^{1}$\thanks{E-mail:
temury@genao.org} and G. Belvedere$^{2}$\thanks{E-mail:
gbelvedere@ct.astro.it}\\
$^{1}$Georgian National Astrophysical Observatory (Abastumani
Astrophysical Observatory), Al. Kazbegi ave. 2a, 0160 Tbilisi,
Georgia\\
$^{2}$Dipartimento di Fisica e Astronomia, Sezione Astrofisica,
Universit{\'a} di Catania, Via S.Sofia 78, I-95123 Catania, Italy}
\begin{document}

\date{}

\pagerange{\pageref{firstpage}--\pageref{lastpage}} \pubyear{2005}

\maketitle

\label{firstpage}

\begin{abstract}
We suggest that the observed long-period Alfv{\'e}n waves in the
solar wind may be generated in the solar interior due to the
pulsation of the Sun in the fundamental radial mode. The period of
this pulsation is about 1 hour. The pulsation causes a periodical
variation of density and large-scale magnetic field, this
affecting the Alfv{\'e}n speed in the solar interior. Consequently
the Alfv{\'e}n waves with the half frequency of pulsation (i.e.
with the double period) can be parametrically amplified in the
interior below the convection zone due to the recently suggested
swing wave-wave interaction. Therefore the amplified Alfv{\'e}n
waves have periods of several hours. The waves can propagate
upwards through the convection zone to the solar atmosphere and
cause the observed long-period Alfv{\'e}n oscillations in the
solar wind.

\end{abstract}


\section{Introduction}

Recent {\it in situ} observations of solar polar regions by {\it
Ulysses} spacecraft have shown the presence of long-period
(periods of several hours) outwardly propagating Alfv{\'e}n waves
\citep{smith,balogh}. The Alfv{\'e}n waves may play a significant
role in solar wind acceleration, therefore it is very important to
understand the process of their generation. As the observed
Alfv{\'e}n waves show predominantly outward propagation, they
probably are of solar origin. Unfortunately no clear physical
mechanism explaining the generation of such long-period Alfv{\'e}n
waves has been suggested so far.

It is an interesting fact, that the low-frequency oscillations
have been observed also in solar lower atmosphere
\citep{merk1,merk2,mash}. The Amplitude of observed oscillations
grows from the center of the solar disc to the limb i.e. the waves
are rather tangential than radial. Therefore the authors suggested
that they can be Alfv{\'e}n waves propagating upwards from the
solar interior. Thus the energy source of these Alfv{\'e}n waves
can be located well below the solar surface.

Here we suppose that the long-period Alfv{\'e}n waves can be
generated in the solar interior by pulsation either in the
fundamental radial mode and/or in low frequency g-modes through
the recently suggested {\it swing wave-wave interaction}
\citep{zaq0,zaq1,zaq2}. The period of the solar fundamental radial
mode is $\sim$ 1 hour; periods of g-modes can be longer. The
presence of a large-scale weak magnetic field in the interior can
not affect the pulsation significantly. However the pulsation
causes the periodical variation of some medium parameters (such as
density and magnetic field) and consequently of the Alfv{\'e}n
speed in the interior. But the periodical variation of Alfv{\'e}n
speed either due to sound waves \citep{zaq0,zaq3} or fast
magnetosonic waves \citep{zaq1} leads to the amplification of
Alfv{\'e}n waves with twice the period of compressible waves. By
this mechanism, it has been shown that the pulsation in the
fundamental mode of solar-like stars in binary systems, which is
excited by the companion's gravity, may lead to the amplification
of double period Alfv{\'e}n waves \citep{zaq2}. Thus the question
naturally arises whether a similar process may occur in the Sun.
In this case, the solar fundamental pulsation may amplify
torsional Alfv{\'e}n waves in the interior with the double period
i.e. $\sim$ 2 hours, while g-modes may amplify Alfv{\'e}n waves
with longer periods.

The generated Alfv{\'e}n waves may propagate upwards and cause the
observed oscillations in the photosphere/chromosphere and the
solar wind. It is a very intriguing fact that one of observed
oscillation period in the chromosphere is $\sim$ 2 hours
\citep{merk1}.

In this letter we show how the solar fundamental radial pulsation
may amplify the torsional Alfv{\'e}n waves in the interior. As to
the coupling of g-modes and Alfv{\'e}n waves, it is more
complicate to describe it mathematically, therefore we will study
it in future.

\section[]{Statement of the problem and developments}

We start from the ideal magnetohydrodynamic (MHD) equations:
\begin{equation}
{{{\partial {\rho}}}\over {\partial t}} +
{\nabla}({\rho}{\bf v})=0,
\end{equation}\begin{equation}
{\rho}{{{\partial \bf v}}\over {\partial t}} +
{\rho}({\bf v}{\cdot}{\nabla})
{\bf v} = - {\nabla}p + {1\over {4\pi}}({\nabla}{\times}{\bf
B}){\times}{\bf B} -
{\rho}{\nabla}{\phi},
\end{equation}
\begin{equation}
{{{\partial \bf B}}\over {\partial
t}}={\nabla}{\times}({\bf v}{\times}{\bf B}),
\end{equation}
\begin{equation}{\nabla}^2{\phi}=4{\pi}G{\rho},
\end{equation}
\begin{equation}
p=p_0\left ({{\rho}\over {\rho_0}}\right )^{\gamma},
\end{equation}
where $\rho$ is the medium density, $p$ is the
pressure, ${\bf v}$
is the velocity, $\bf B$ is the magnetic field, G is the gravitational constant, $\phi$ is the gravitational potential and $\gamma$ is the ratio of specific heats. We consider a medium with zero viscosity, infinite
conductivity and negligible displacement current.
We argue that the coupling between pulsation and Alfv{\'e}n
waves occurs below the convection zone, therefore we do not consider convective effects
and consequently the $\alpha$-effect term in the Mean
Field Electrodynamics induction equation (see e.g. Belvedere 1985) which represents a mean
electromotive force generated by cyclonic convection. Here and in the remaining part of the paper we also
neglect the rotational effects. 

\subsection[]{Fundamental pulsation of the Sun}

The Sun is considered to be in hydrostatic equilibrium so that its
own gravity is balanced by the pressure gradient.

A slight deviation from the equilibrium leads to an oscillation
which may be studied by the linear perturbation theory. Stellar
oscillations (in the absence of magnetic field) can be divided
into two classes: high-frequency p-modes or pressure modes, whose
restoring force is the pressure gradient, and low-frequency
g-modes, where buoyancy is the restoring force. There are also
intermediate oscillations sometimes called f-modes \citep{cow41}.

\begin{figure}
 \centering \includegraphics[width=8cm]{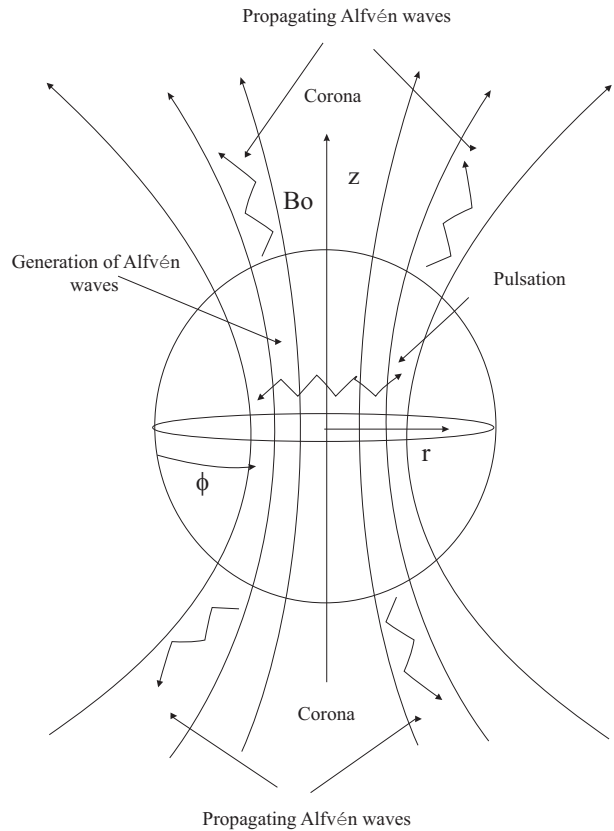}
 \caption{Schematic picture of
generation and propagation of low frequency Alfv{\'e}n waves in
the Sun. The Alfv{\'e}n waves are generated in the solar interior
due to the pulsation of the Sun in the fundamental radial mode.
Then they propagate along magnetic field lines towards the
corona.}
\end{figure}

The boundedness of any system usually leads to oscillation in the
first eigenfrequency which is also called the fundamental
frequency of the system. The mechanical analog of the fundamental
oscillation is the tuning fork in the case of an impulsive force
and the musical trumpet in the case of a continuous force. The
oscillation of a star in the fundamental mode yields a wavelength
comparable with the stellar radius. Therefore the fundamental
frequency evaluated for an homogeneous self-gravitating sphere
will be of order of ${\sqrt {GM/R^3}}{\sim}c_s/R$, where  $R$ and
$M$ are the radius and the mass of star and 
$c_s=\sqrt{{\gamma}p_0/{\rho}_0}$ is the mean sound
speed in the interior.
However the real equilibrium density is not uniform throughout the
Sun, therefore the fundamental period is shorter $\sim$ 1 hour.
Unfortunately no clear observational evidence of the fundamental
pulsation was found, which is probably due to the acoustic cutoff
in the stratified atmosphere. The acoustic cutoff frequency in an
isothermal atmosphere is $c_s/2{\Lambda}_0$, where ${\Lambda}_0$
is pressure scale height. Under conditions of the
photosphere/chromosphere ${\Lambda}_0 \approx 125$ km. Then for
the typical photospheric sound speed $c_s=7.5$ km/s the cutoff
frequency is 0.03 s$^{-1}$, which gives the cutoff period of 210
s. This means that the sound waves with longer periods, such as
the fundamental radial pressure mode, can not pass through the
photosphere. Probably this is a reason why the low-frequency
pressure modes are not seen in oscillation spectrum (but see a
possible evidence of low frequency pulsation in \citep{brown}).

As spherical coordinates lead to a more complicated mathematical
formalism, we adopt a cylindrical coordinate system ($r$, $\phi$,
$z$) and for simplicity consider only the axisymmetric problem so
everywhere ${{\partial }/{\partial {\phi}}}=0$ is assumed. We
suppose that an uniform poloidal magnetic field, $B_0$, exists
inside the star directed along the $z$ axis (see Fig.1). The
large-scale magnetic field is generated either by classical dynamo
or other mechanisms, but this is beyond the scope of the present
paper. We just assume that the magnetic field exists. This field
is considered to be relatively weak, so it does not affect the
fundamental pulsation significantly.

In the adopted cylindrical coordinate system, we approximate the
fundamental solar oscillation as a radial pulsation of the
cylinder, thus neglect the motion along the $z$ axis. Then the linearised 
equations (1)-(5) can be split into radial and azimuthal components, where 
the radial component corresponds to pulsations and the azimuthal component 
to torsional Alfv{\'e}n waves. Torsional Alfv{\'e}n waves have azimuthal 
velocity polarisation and propagate along the $z$ axis. As we will see later,
the wavelength of the Alfv{\'e}n waves is much smaller as compared to the solar radius 
(due to small Alfv{\'e}n speed in the interior), therefere the $z$ velocity component 
of the global pulsation, which has a spatial scale comparable to the solar radius, can be 
considered as homogeneous in Alfv{\'e}n spatial scale and therefore cannot significantly 
affect the Alfv{\'e}n wave dynamics. 
Therefore, the motion along the $z$ axis due to the global pulsation may only 
complicate the calculations and thus is neglected as mentioned above.      

The radial part of linearised equations (1)-(5) in cylindrical coordinates is

\begin{equation} {{{\partial b_{z}}}\over
{\partial t}}=
-B_{0}{{{\partial u_r}}\over {\partial r}} -
{{{\partial
B_{0}}}\over {\partial r}}u_r -
{{B_{0}}\over r}u_r,
\end{equation}\begin{equation}
{{{\partial {\delta}{\rho}}}\over {\partial t}}=
-\rho_0{{{\partial
u_r}}\over {\partial r}} -
{{{\partial {\rho_0}}}\over {\partial r}}u_r -
{{\rho_0}\over r}u_r,
\end{equation}\begin{equation}
{{\rho_0}{{{\partial u_r}}\over {\partial t}} = -
{{{\partial}}\over
{\partial r}}\left[{\delta}p +
{{B_{0}b_{z}}\over
{{4\pi}}}\right ] - {\delta
{\rho}}{{{\partial {{\phi}}}}\over {\partial
r}}} , \end{equation}
\begin{equation}{{{\partial {\delta}p}}\over {\partial
t}} + u_r{{{\partial
{p_0}}}\over {\partial r}}= c_s^2\left ({{{\partial
{\delta}{\rho}}}\over
{\partial t}} + u_r{{{\partial {{\rho}_0}}}\over
{\partial r}} \right ),
\end{equation}
while the azimuthal part is   
\begin{equation}
{{\partial b_{\phi}}\over {\partial t}}=  B_0{{\partial
u_{\phi}}\over {\partial z}},
\end{equation}
\begin{equation} {{\partial
u_{\phi}}\over {\partial t}}={{B_0}\over {4\pi{\rho_0}}}{{\partial
b_{\phi}}\over {\partial z}},
\end{equation}
where $u_r, u_{\phi}, b_{z}, b_{\phi}, {\delta}{\rho}$
and ${\delta}p$ are the
velocity, magnetic field, density and
pressure perturbations respectively.
Equations (6)-(9) govern the linear radial pulsation of
the star, while
equations (10)-(11) describe the linear torsional Alfv{\'e}n
waves. 

In general, stellar radial adiabatic pulsations can be represented
as standing spherical waves \citep{cox} with a linear radial
velocity field
\begin{equation}u_r={\alpha}F(r){\sin}(\omega_nt),
\end{equation}
where $\omega_n$ is the eigenfrequency, $F(r)$ is the
eigenfunction and $\alpha$ is the pulsation amplitude. The
expression of the eigenfunction $F(r)$ depends on the spatial
profiles of the unperturbed physical quantities and may be
represented by some combination of spherical Bessel functions.
However, here we are not interested in specific pulsation
functional forms, therefore we retain its general form $F(r)$,
which then may be specified by choosing the spatial distribution
of the unperturbed quantities throughout the Sun.

Then, using
expression (12), the linearised continuity and induction equations
give the density and magnetic field perturbations as:
\begin{equation}
\delta\rho={{\alpha}\over
{\omega_n}}{\cos}(\omega_nt)F_{\rho}(r),\end{equation}
\begin{equation}
b_z={{\alpha}\over
{\omega_n}}{\cos}(\omega_nt)F_{b}(r),\end{equation}
where\begin{equation} F_{\rho}(r)={\rho_0(r)}{{\partial F(r)}\over
{\partial r}} + \left ({{\partial {\rho_0(r)}}\over {\partial r}}
+ {{{\rho_0(r)}}\over r}\right
)F(r),\end{equation}\begin{equation} F_{b}(r)= B_0(r){{\partial
F(r)}\over {\partial r}} + \left ({{\partial {B_0(r)}}\over
{\partial r}} + {{{B_0(r)}}\over r}\right )F(r),\end{equation}
where $\rho_0(r)$ is the unperturbed density.

Equations (13)-(14) show that the radial pulsation leads to local
periodical variation of density and magnetic field throughout the
Sun, which causes the periodical variation of the Alfv{\'e}n speed
at each level $r$. In next section we show that it determines an
exponential amplification of torsional Alfv{\'e}n waves with half
the frequency of pulsation.

\subsection[]{Resonant torsional Alfv{\'e}n waves in the solar interior}

The radial pulsation affects the torsional Alfv{\'e}n waves
through nonlinear interaction and then equations (10)-(11) can be rewritten
as
\begin{equation}
{{\partial b_{\phi}}\over {\partial t}}=  (B_0 + b_z){{\partial
u_{\phi}}\over {\partial z}} - {{\partial (u_rb_{\phi})}\over
{\partial r}},
\end{equation}
\begin{equation}
{{\partial u_{\phi}}\over {\partial t}} + {{u_{r}}\over
{r}}{{\partial (ru_{\phi})}\over {\partial r}}={{(B_0 + b_z)}\over
{4\pi({\rho_0} + \delta\rho)}}{{\partial b_{\phi}}\over {\partial
z}}.
\end{equation}

Thus equations (17)-(18) describe the evolution of Alfv{\'e}n waves in presence of the global pulsation.
Since the energy associated to this pulsation is clearly much larger than the energy stored in Alfv{\'e}n waves,
the back reaction of Alfv{\'e}n waves can be ignored (at least at the initial phase of the wave evolution), 
which means that the amplitude of the pulsation remains constant.

Equations (17)-(18) can be further simplified by taking the
Alfv{\'e}n oscillations in narrow cylindrical shells, then the
unperturbed density $\rho_0$ and magnetic field $B_0$ can be
considered as homogeneous (thus $r$ will stand as a parameter in the equations) and 
equations (13)-(14) give

\begin{equation}
{{\delta\rho}\over {{\rho}_0}}={{b_z}\over {B_0}}.
\end{equation}

Then equations (17)-(18) lead to the second order differential
equation
$$
{{\partial^2 u_{\phi}}\over {\partial t^2}} + \left [{1\over
r}{{\partial (ru_r)}\over {\partial r}}\right ]{{\partial
u_{\phi}}\over {\partial t}} - \left [{{B_0(B_0 + b_z)}\over
{4\pi{\rho_0}}}{{\partial^2 u_{\phi}}\over {\partial z^2}} -
{{u_{\phi}}\over {r}}{{\partial u_r}\over {\partial t}} \right ]=
$$
\begin{equation}
=0.
\end{equation}

Using the Fourier transform of $u_{\phi}$ with $z$-dependence
\begin{equation}
u_{\phi}=\int{{\hat u}_{\phi}e^{ik_zz}dk_z},
\end{equation}
equation (20) is rewritten as

$$
{{\partial^2 {\hat u}_{\phi}}\over {\partial t^2}} + \left [{1\over
r}{{\partial (ru_r)}\over {\partial r}}\right ]{{\partial
{\hat u}_{\phi}}\over {\partial t}} + \left [{{B_0(B_0 + b_z)}\over
{4\pi{\rho_0}}}k^2_z +
{{1}\over {r}}{{\partial u_r}\over {\partial t}} \right ]{\hat u}_{\phi}=
$$
\begin{equation}
=0.
\end{equation}

Now, using the well known substitution of function 
\begin{equation}
{\hat u}_{\phi}={\tilde u}_{\phi}e^{-{1\over {2}}\int{{{{\partial (ru_r)}\over {r\partial
r}}}dt}},
\end{equation}
the term with the first time derivative can be removed and with expressions (12)-(14) we finally get 
\begin{equation}
{{\partial^2 {\tilde u}_{\phi}}\over {\partial t^2}} + \left
[v^2_Ak^2_z + {{\alpha{\Psi}(r)}\over {\omega_n}}{\cos}(\omega_nt)
\right ]{\tilde u}_{\phi}=0,
\end{equation}
where
\begin{equation}
\Psi(r)=\left [v^2_Ak^2_z - {{\omega^2_n}\over 2} \right
]{{\partial u_r}\over {\partial r}} + \left [v^2_Ak^2_z +
{{\omega^2_n}\over 2} \right ]{{u_r}\over r} .
\end{equation}
Without the pulsation (i.e. taking $\alpha$=0), equation (24) describes 
the usual dispersion relation of Alfv{\'e}n waves. But, in presence of the forcing introduced by the  
pulsation ($\alpha {\not =} 0$), equation (24) is the well known Mathieu equation. 
The solution of this equation with frequency ${\omega_n}/2$ has an exponentially growing character 
\citep{lan}, thus the main resonant solution occurs when \citep{zaq1,zaq2}
\begin{equation}{\omega_A}= v_Ak_z {\approx}{{\omega_n}\over 2},\end{equation}
where $\omega_A$ is the frequency of torsional Alfv{\'e}n waves.
Under condition (26) the solution of equation (24) is 
\begin{equation}
{\tilde u}_{\phi}(t)={\tilde u}_0e^{{{\left
|{\alpha}{\Psi(r)}\right |}\over {2\omega_n}}t}\left
[{\cos}{{\omega_n}\over 2}t {\mp} {\sin}{{\omega_n}\over 2}t
\right ],
\end{equation}
where ${\tilde u}_0={\tilde u}(0)$ and the phase sign depends on
${\alpha}{\Psi(r)}$; it is $+$ for negative ${\alpha}{\Psi(r)}$
and $-$ for positive ${\alpha}{\Psi(r)}$. Note that the solution
has a resonant character within the frequency interval

\begin{equation} {\left |{\omega_A} -
{{\omega_n}\over 2} \right |}<{\left |{{\alpha\Psi}\over
{\omega_n}} \right |}.
\end{equation}

Thus the radial pulsation leads to an exponential amplification of
torsional Alfv{\'e}n waves with the half frequency ${1\over
2}{\omega_n}$. The growth rate depends on the amplitude $\alpha$
and spatial structure of the pulsation eigenfunction $F(r)$, which
in turn depends on the radial structure of density $\rho_0$ and
magnetic field $B_0$. The resonant condition (26) imposes a
restriction on the wave number $k_z$. In other words, the
pulsation "picks up" the harmonics of torsional Alfv{\'e}n waves
with a certain wavelength at each radial distance $r$.

Numerical simulation of equations (17)-(18) also shows the resonant
amplification of torsional Alfv{\'e}n waves. For simplicity we
take the function $F$ in expression (12) as constant and look for
the temporal evolution of the spatial Fourier harmonics of
$b_{\phi}$ and $u_{\phi}$. The results are shown on Fig. 2. The
exponential amplification of the Alfv{\'e}n wave spatial Fourier
harmonics (upper plots) with the double period of pulsation is
clearly seen.

\begin{figure}
 \centering \includegraphics[width=9cm]{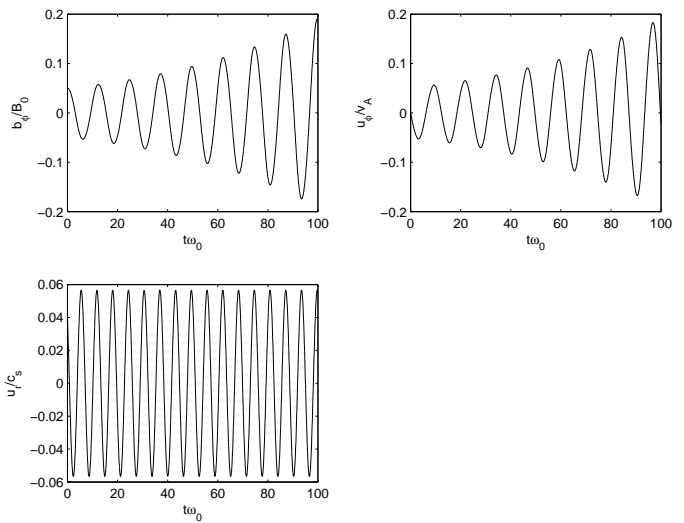}
 \caption{Resonant amplification of
Alfv{\'e}n waves due to the radial pulsation. The plot below is
the velocity of the fundamental pulsation. The plots above are the
magnetic field and the velocity components of the torsional
Alfv{\'e}n waves. These waves have the double period of pulsation.
The energy of pulsation is considered to be much larger than the
energy of the Alfv{\'e}n waves, therefore the amplitude of
pulsation remains constant.}
\end{figure}

As the main resonance occurs at the half frequency of pulsation
(see equation (26)), the resonant Alfv{\'e}n waves in the solar
interior will have a period
\begin{equation}
T_A = 2T_0 \approx 2 \,\, hours,
\end{equation}
where  $T_0 \approx 1$ hour is the pulsation period.

Due to the resonant range (28), the generated Alfv{\'e}n waves
will have not a fixed period, but many periods in that range. The
width of range depends on the amplitude of pulsation. Thus the
period of generated Alfv{\'e}n waves in the solar interior may be
in the range of
\begin{equation}
T_A \sim 1.5-3\,\, hours.
\end{equation}

From expression (26) we may calculate the wavelength of the
generated Alfv{\'e}n waves. Using the dispersion relation of the
sound waves, we have $\lambda / R \sim {{v_A}/{c_s}}$, where
$\lambda$ is the wavelength of the resonant Alfv{\'e}n waves and
$v_A=B_0/\sqrt{4\pi\rho_0}$ is the mean Alfv{\'e}n speed in the
interior. As the plasma $\beta$ ($\beta \approx c^2_s/v^2_A$) is
higher in the solar interior, the wavelength of the Alfv{\'e}n
waves will be shorter comparing to the solar radius. Taking
$c_s/v_A=100$, we get a value of this wavelength in the solar
interior such as $\lambda \sim 7\,\,000\,\, km$.

It must be mentioned that the amplification of the Alfv{\'e}n
waves due to the fundamental pulsation can occur in the solar
interior below the convection zone, as the motions in the
convection zone may destroy the resonance process.

The amplified Alfv{\'e}n waves can propagate along the magnetic
field lines through the convection zone and reach the solar
atmosphere (see the schematic picture in Fig. 1). The
configuration of large-scale magnetic field is very simplified as
the complication due to the convection is completely ignored. In
fact the stochastic small scale magnetic fields in the convection
zone may make the wave propagation difficult. But the magnetic
fields in sunspots and in regions between large convective cells
are predominantly large scale and they may easily guide the waves
through the convection zone. Therefore the Alfv{\'e}n waves
probably can be observed in sunspot latitudes and in polar regions
where the dipole component of magnetic field dominates. It is an
interesting fact that observed low-frequency oscillations found in
photosphere, chromosphere and prominence by \citet{mash} have the
period of $\sim 1.75$ hours at sunspot (25$^0$-30$^0$) and higher
(60$^0$) latitudes. The observed oscillations are rather
transversal than the radial and can be associated to the
Alfv{\'e}n waves propagating from the solar interior. Also
long-period transversal oscillations with the period of $\sim 2$
hours have been observed in chromospheric spectral lines
\citep{merk1,merk2} and again can be interpreted as Alfv{\'e}n
waves. The long-period oscillations recently observed in solar prominence 
\citep{jaum} can be also due to the Alfv{\'e}n waves propagating from below.
It is also intriguing fact that long-period Alfv{\'e}n
waves are observed in fast solar wind at high latitudes where
dipolar magnetic field component is dominant \citep{smith,balogh}.
It is quiet possible that photospheric and chromospheric
Alfv{\'e}n waves may propagate into the corona and solar wind. The
amplitude of long-period oscillations \citep{mash} in the
photosphere is $\sim 0.1$ km/s and in the chromosphere ${\sim
0.25}$ km/s. If these oscillations propagate into the low density
corona then the amplitudes must be increased as the wave energy
density remains approximately constant. The wave energy density is
of order $\sim \rho u^2$, where $\rho$ is the background density
and $u$ is the wave velocity. The particle density in the
photosphere is 10$^8$ times more than in the corona. Thus in order
to retain the constant energy density, the wave velocity must
increase 10$^4$ times i.e. to the value $\sim 10^3$ km/s. Which
means that the waves become intrinsically non-linear as the
Alfv{\'e}n speed has the similar value in the corona. It is the
fact that solar wind Alfv{\'e}n waves are nonlinear ${\delta V}/V
\sim {\delta B}/B \sim 1$.

Another question is the energy balance between solar global modes
and observed Alfv{\'e}n waves. The energy stored in the
fundamental pulsation is huge as the whole Sun takes part in the
oscillation. Also the ratio between the densities in the interior
(say the base of convection zone) and the photosphere is $\sim
10^7$. Therefore even low amplitude pulsation can be responsible
for observed Alfv{\'e}n wave energy.

On another hand, as we mentioned above there is no clear
observational evidence of solar global pulsation. \citet{brown}
reported the variation of solar diameter with the periods around 1
hour. But since the pulsation was not clearly observed in solar
oscillation spectrum. There can be two main reasons why the
fundamental pressure mode is absent at the photosphere. Either it
has very low amplitude or it can not pass the photosphere due to
the stratification cutoff (see above). On another hand, the
continuous energy transfer from the pulsation into the Alfv{\'e}n
waves in the interior may significantly lower the amplitude of
pulsation at the surface.

As the mean Alfv{\'e}n speed varies from the solar interior to the
corona, the wavelength and the period of these upwards propagating
Alfv{\'e}n waves will change along their way to the corona.
Therefore the observed Alfv{\'e}n oscillations in the solar wind
will have a longer wavelength than the Alfv{\'e}n waves in the
interior.

It must be mentioned that many simplifications (such as the cylindrical 
symmetry instead of the spherical one, the spatially homogeneous distribution of 
physical quantities) have been made to outline the physics of the 
Alfv{\'e}n wave generation in a simple way. Of course, the real situation is much more complicated, 
but the physical meaning of the mechanism will remain the same. Future  exstensive numerical simulation is needed in order to take into account the real distribution of physical 
quantities, but this is beyond the scope of this letter.      

\section{Conclusions}

We suggest that outwardly propagating long-period Alfv{\'e}n
waves, which have been observed in solar polar regions by {\it
Ulysses} \citep{smith,balogh} can be generated in the solar
interior due to the solar fundamental radial pulsation. It is
shown that this radial pulsation parametrically amplifies
torsional Alfv{\'e}n waves. The main resonance occurs for
Alfv{\'e}n waves with the double period of pulsation, although the
process has a resonant range of frequencies, whose width depends
on the amplitude of pulsation. Therefore the solar fundamental
radial pulsation, with a period of $\sim$ 1 hours, will amplify
Alfv{\'e}n waves with a period of $\sim$ 1.5-3 hours. These
amplified Alfv{\'e}n waves may propagate upwards to the solar
corona and cause the observed magnetic field and velocity
oscillations in the solar wind.

It must be mentioned that low-frequency g-modes may also amplify
Alfv{\'e}n waves with longer periods, but we do not consider this
process in this letter. However it is really worth to study the
process in future, as the existence of g-modes in the solar
interior could be revealed through Alfv{\'e}n oscillations in the
photosphere/chromosphere and solar wind.

\section*{Acknowledgements}

The work of T.Z. was supported by the NATO Reintegration Grant
FEL.RIG 980755 and a contribution of
the Sezione Astrofisica del Dipartimento di Fisica e Astronomia
dell'Universit{\'a} di Catania. We thank an anonymous referee for careful interest and useful
suggestions to improve the
paper.

\label{lastpage}

\end{document}